\documentclass[a4paper, 12pt]{article}

\usepackage[a4paper]{geometry}

\title{Thermodynamics and microscopic theory: \\[0.1cm]  An educational proposal for the high school}

\author{A.~Ercoli  \\[0.1cm] {\small Liceo Scientifico Statale "P. Ruffini" di Viterbo - Viterbo, Italy} 
\\[0.3cm]
V.~Lubicz \\[0.1cm] {\small Dipartimento di Matematica e Fisica, Universit\`a Roma Tre - Roma, Italy}}

\date{}

\begin{document}

\maketitle

\begin{abstract}
We present an educational proposal which aims to illustrate the elegant, refined and coherent physics contained in Thermodynamics, through a path which assigns to the microscopic description of the physical systems a constantly privileged role. This approach allows to reach a simple and, at the same time, deep understanding of the laws of Thermodynamics, while still emphasizing their great generality, which permits their application to all macroscopic systems, from simple gases to black holes, arriving to characterize the evolution of the entire Universe.
\end{abstract}

\section{Introduction}
The history of physics shows us that Thermodynamics has always provided a solid reference for the theoretical developments, even and especially during the major advances in scientific theories. For instance, in the historical and pioneering papers by Planck on the black body radiation and by Einstein on the photoelectric effect and on the Brownian motion the arguments are based precisely on purely thermodynamic considerations, whose laws have such a broad generality that can be applied in the most different contexts.

In spite of that, the teaching of Thermodynamics in Italian high school is getting increasingly reduced in recent years. One also finds that in many cases, at the end of the studies, students have reached quite a limited view of Thermodynamics, which too often seems to concern only ideal gases and heat engines. In this way, they loose the possibility of appreciating a theory which could provide, instead, completeness and new perspectives to what they have previously learned.

It is also not so rare to find that the physical contents of Thermodynamics are poorly understood by the students, with concepts that are often unclear if not, in some cases, even erroneous. A typical difficulty, for example, consists in properly distinguishing the quantities {\em heat}, {\em work} and {\em energy}. Heat and work are not forms of energy, which, as such, are possessed by the physical system. Rather, they are ways through which energy is exchanged. This concept, if not properly assimilated, can easily hinder the understanding of the first law of Thermodynamics and of the principle of energy conservation.

Another difficulty, concerns the definition of entropy. While the physical meaning of entropy, as explained to us by Boltzmann, is eventually understood by the students, it remains often obscure how the number of microscopic states is related to the Clausius definition of entropy, i.e. to the ratio $\Delta Q/T$ between exchanged heat and temperature in a transformation, and why this ratio does never decrease for an isolated system.

 From an educational point of view, the presentation of the second law based on the two historical statements by Clausius and Kelvin is quite unsatisfactory. Besides the rather artificial proof of the equivalence of the two statements, what is their deep physical meaning, that also explains, in turn, their equivalence? In addition, both the Clausius and Kelvin statements, as well as the Carnot's theorem, are strongly linked to the historical context in which they have been derived. In this way, the idea is transmitted to the students that the second principle of Thermodynamics mainly concerns the efficiency of heat machines, without showing the vast generality of this law which establishes, for any physical process, the unique direction in time in which it can occur.

 Starting from these considerations, we developed an educational proposal, published in a textbook (in Italian)~\cite{libro}, which aims to illustrate the elegant, refined and coherent physics contained in Thermodynamics. 
 
 The conceptual strength of Thermodynamics, and at the same time its simplicity, stems from the great generality of its physical laws, whose derivation can disregard completely the knowledge of the microscopic structure of the physical systems. We believe, however, that a real, deep and even simpler understanding of the laws of Thermodynamics can only be achieved by starting from the microscopic description of the systems. This belief has led us to assign to the microscopic theory in our proposal a constantly privileged role\footnote{An educational approach, devoted to undergraduate university courses, which shares with us a similar point of view, is described in~\cite{Reif}. See also \cite{Malgieri} for a teaching learning sequence for the high school mainly relying on the microscopic approach.}. It is important, at the same time, that students are able to appreciate the great generality of the thermodynamic laws, which do not concern only ideal gases or heat machines, but apply to all macroscopic physical systems, from gases to black holes, arriving to characterize the evolution of the entire Universe.

\section{Some illustrative examples}
In this section, in order to present to the reader a more concrete idea of our proposal, we briefly discuss some of those which are, in our opinion, its most significant contents.

As already emphasized, the most characterizing aspect of our proposal is the assignment to the microscopic theory of a constantly privileged role. For the explanation of the main concepts, such as work or heat, as well as for several of the thermodynamic processes which are discussed, a microscopic description is also provided. Although this description definitively requires more details than the macroscopic treatment, it is nevertheless simple and clear for the students, since it only requires the application of the laws of motion, the knowledge of the general principles of mechanics and simple probabilistic arguments. As noted in~\cite{Besson}, ``the explanations proposed in Thermodynamics can be unsatisfactory for the student need of understanding, since they often only show how things should or should not be, but not how things actually happen".

\subsection{The microscopic treatment of work and heat}
Work is the energy exchanged between two systems when a displacement occurs under the action of a force. From the very definition of work, as the product of the force times the displacement, it follows that it is always the work done on the system by the {\em external forces} (rather than the work done by the system) to be responsible for the variation of the kinetic energy of the molecules, and therefore of the temperature of the system. We find that this point is often not enough emphasized in many thermodynamics textbooks.

Heat instead is the transfer of energy between two systems due to a difference in temperature. Since in turn the temperature is a measure of the thermal agitation of the molecules, one can equivalently state that heat is the transfer of energy between two systems due to the work done by the microscopic forces exerted in the collisions among the molecules in thermal agitation. We are thus also led to the conclusion that, at the microscopic level, the distinction between heat and work vanishes.

As an example of microscopic treatment of work in our textbook, we discuss here the case of the adiabatic expansion, or compression, of an ideal gas. When reversible, this transformation is described at the thermodynamic level by the Poisson equations
\begin{equation}
\label{Poisson}
    p\, V^\gamma = costant \qquad \mathrm{or} \qquad T\, V^{\gamma-1} = costant \ \ ,
\end{equation} 
which relate pressure, volume and temperature of the gas during the transformation.

The thermodynamic derivation of Eq.\,(\ref{Poisson}), based on the first law, is straightforward. However, it is only with the microscopic description that one arrives at a real understanding of the physical process. By using the laws of energy and momentum conservation in the collision between a molecule of the gas and the moving piston, we can compute the change of velocity of the molecule. One finds, as expected, that the molecule slows down in the expansion, when the piston moves away from it, while its speed increases in the compression, when the piston approaches to it. Starting from this velocity change of the molecule, one can compute the variation of the average kinetic energy of the whole gas and, therefore, of its temperature. This detailed, but physically intuitive discussion, leads directly to the Poisson equations (\ref{Poisson}).

\subsection{The microscopic definition of entropy}
A clear understanding of the intrinsic irreversibility we observe in nature for most of the physical processes, and their occurrence in only one direction in time (the egg becomes omelette, but the omelette never turns into an egg!) can only be reached through the microscopic description. As explained by Boltzmann, all physical processes occur in the direction in which macrostates with higher and higher probability are subsequently reached, corresponding to an increasing number of microscopic states.

Motivated by this observation, in our textbook we introduce entropy starting from the microscopic Boltzmann definition, $S=k \log W$, preceded and followed by a number of examples and applications. It is only later, in the next chapter, that the Clausius thermodynamic definition $\Delta S=\Delta Q_{rev}/T$ is presented. In this respect, therefore, we do not follow the historical order which is followed, instead, by the majority of textbooks.

A quite critical aspect in understanding the entropy concerns the equivalence of the Boltzmann and Clausius definitions, which look so different one from each other. Why ``$S=k \log W$" is equivalent to ``$\Delta S=\Delta Q_{rev}/T$"? In our textbook, we provide an argument for this equivalence for the ideal gas, based on the counting of microscopic states. To the best of our knowledge, the issue is never addressed in textbooks for the high school (and even in the introductory physics courses at the undergraduate level).

\subsection{The Clausius inequality and the second law of Thermodynamics}
Historically, among the various (potentially infinite) statements of the second law, two have had particular importance, in connection in particular with the study of heat engines: the Clausius and the Kelvin-Planck statements. As is known, the two statements are equivalent, in the sense that each one can be derived from the other. But the standard proof of this equivalence, when presented, looks rather involved and not intuitive.

In our proposal we start from the observation that the second law can be always expressed in terms of the Clausius inequality,
\begin{equation}
\oint \delta Q/T \leq 0 \ ,     
\end{equation}
which represents therefore its general mathematical formulation. Both the Clausius and the Kelvin-Planck statements, as well as the Carnot's theorem, are thus be derived in our textbook starting from the Clausius inequality. Besides its greater simplicity, this approach has the remarkable advantage of providing a highly unified picture of the second principle of Thermodynamics, one of the physical laws of the greatest impact in our everyday life.

\section{The educational proposal}
As already mentioned, the educational path we are proposing has been published in the form of a textbook (in Italian), entitled {\em Thermodynamics and Microscopic Theory}~\cite{libro}. The book is intended for high school students, but we believe that a similar educational path can be also adapted to be taught in undergraduate physics courses, where the more advanced mathematical skills possessed by the students can be profitably used, making the presentation of the various arguments even more effective.

After an {\em Introduction}, the book contains five chapters entitled: 1. {\em Thermodynamics and kinetic theory}; 2. {\em Energy transfers: work and heat}; 3. {\em Energy and the first law of Thermodynamics}; 4. {\em Entropy and probability}; 5. {\em Entropy and the second law of Thermodynamics}; and it is eventually concluded by an ending section entitled {\em Past, future and the entropy of the Universe}.

A series of exercises are proposed at the end of each chapter, which mainly aim at testing the level of understanding reached by the students. In most of these exercises calculations play a marginal role, while correct arguments and good theoretical understanding are required for their resolution. In this spirit, for each exercise, ideas and suggestions for its resolution are provided at the end of the volume.

\section*{Acknowledgments}
The authors warmly thank the research group in Educational Physics of the Roma Tre University and, in particular, Ilaria De Angelis and Adriana Postiglione, for their precious and continuous support in the development of this project.

\end{document}